\documentclass[11pt]{article}
\usepackage[a4paper,margin=2.3cm]{geometry}
\usepackage{amsmath,amssymb,amsfonts,bm}
\usepackage{graphicx}
\usepackage{booktabs}
\usepackage{enumitem}
\usepackage{listings}
\usepackage{xcolor}
\usepackage{natbib}
\usepackage{hyperref}

\hypersetup{colorlinks=true,linkcolor=blue,citecolor=blue,urlcolor=blue}
\DeclareUnicodeCharacter{00A0}{\space} 
\DeclareUnicodeCharacter{202F}{\space} 
\DeclareUnicodeCharacter{2009}{\space} 
\DeclareUnicodeCharacter{200A}{\space} 
\DeclareUnicodeCharacter{205F}{\space} 
\DeclareUnicodeCharacter{2007}{\space} 
\DeclareUnicodeCharacter{FEFF}{\space} 
\DeclareUnicodeCharacter{200B}{}       
\DeclareUnicodeCharacter{2060}{}       

\newcommand{\rev}[1]{\textcolor{black}{#1}}

\begin{document}

\title{\textsc{Binary‑Star Evolution Modules in
         \textnormal{REBOUNDx}}}
\author{Mohamad Ali-Dib\\
\footnotesize Center for Astrophysics and Space Science (CASS), New York University Abu Dhabi, PO Box 129188, Abu Dhabi, UAE\\
\footnotesize Email: malidib@nyu.edu
}

\date{}
\maketitle
\vspace*{-1.5em}

\begin{abstract}
Close-binary evolution couples Roche–lobe overflow (RLOF), common-envelope (CE) drag, stellar winds, magnetic braking, and gravitational-wave losses, exchanging mass and angular momentum while reshaping orbits and spins. We present interoperable effects in the \textsc{REBOUNDx} extension to \textsc{REBOUND} that embed these processes within high-accuracy $N$-body dynamics. The suite includes: a momentum-conserving RLOF operator with conservative and systemic channels and configurable specific-$j$ loss; a CE drag model based on Mach-dependent dynamical friction with kick limiting; isotropic Reimers winds, Parker-type thermal winds, and Eddington-limited outflows powered by a parametric stellar-evolution module supplying mass-dependent $R$ and $L$; magnetic braking via the Verbunt–Zwaan/Kawaler torque with a saturation-aware closed-form spin update; and post-Newtonian corrections (\rev{2\,PN} point-mass and spin–spin; \rev{2.5\,PN} radiation reaction). Linear momentum is conserved for conservative transfer, a minimal corrective torque enforces angular-momentum consistency, and adaptive sub-stepping stabilizes evolution near contact. Inter-module flags coordinate wind/RLOF/CE activity. The unit-agnostic framework enables self-consistent, time-resolved studies of close binaries in isolated or dynamically rich settings. {Multiple examples and comparisons against other codes are provided in the Appendix.} The code is available at \hyperlink{https://github.com/malidib/ReboundS}{https://github.com/malidib/ReboundS} .
\end{abstract}


\section{Introduction}\label{sec:intro}
Close binary stars evolve under a tightly coupled set of physical processes that redistribute mass and angular momentum and, in some regimes, remove orbital energy. Mass exchange through Roche–lobe overflow (RLOF) is triggered once a donor fills its volume–equivalent Roche surface, typically parameterized by the Eggleton formula \citep{Eggleton1983}, while the instantaneous mass–loss response near the inner Lagrange point can be captured by exponential prescriptions calibrated to the photospheric scale height \citep{Ritter1988}. In more extreme phases, unstable mass transfer and orbital shrinkage may lead to common–envelope (CE) evolution, during which hydrodynamic drag within an extended envelope extracts orbital energy and angular momentum \citep{Ostriker1999}. Independent channels of mass and angular–momentum loss operate throughout a binary’s life: cool convective stars experience magnetic braking via magnetized winds \citep{Verbunt1981,Kawaler1988}, giant–branch stars lose mass through stellar winds that correlate with global parameters \citep{Reimers1975}, and compact binaries emit gravitational radiation that drives secular inspiral on relativistic timescales \citep{Peters1964}. Changes in stellar structure modulate these pathways by altering radii, luminosities, and internal moments of inertia \citep{Hurley2002}. Taken together, these effects govern whether a close binary undergoes stable mass exchange, shrinks into contact and merges, or survives to produce compact remnants.

Capturing this multi-physics landscape with a single method remains challenging. Fully three dimensional hydrodynamic calculations can resolve RLOF streams and CE flows, but they are too expensive for long-term evolution or population-scale parameter studies. Conversely, rapid binary–evolution formalisms encode RLOF, winds, magnetic braking, and gravitational–wave losses through secular prescriptions that are efficient and predictive in isolation, yet they generally assume a two-body context and cannot natively account for higher–order gravitational dynamics (e.g., triples, resonant encounters, or cluster potentials) that can modulate or even trigger mass transfer. A complementary approach is to embed vetted secular and dissipative processes within a high-accuracy $N$-body integrator so that orbital dynamics, spin evolution, and mass exchange proceed self-consistently on a common timestep while preserving the bookkeeping demanded by conservation laws.

This paper presents a suite of new binary–evolution modules implemented in the \textsc{REBOUNDx} extension to the \textsc{REBOUND} $N$-body framework \citep{ReinLiu2012,reintamayo,Tamayo2020}. The modules target the dominant channels relevant to close binaries: (i) a momentum–conserving RLOF and CE module that combines conservative transfer with non-conservative systemic mass loss and supports multiple specific angular momentum prescriptions for escaping gas; (ii) a magnetic–braking operator implementing the Verbunt–Zwaan/Kawaler torque with a saturation branch appropriate for rapidly rotating convective envelopes \citep{Verbunt1981,Kawaler1988}; (iii) isotropic stellar–wind mass loss following the Reimers scaling \citep{Reimers1975}; (iv) thermally driven (Parker-like) winds parameterized by heating efficiency; and (v) post-Newtonian (PN) corrections providing conservative 2\,PN point–mass and spin–spin terms and 2.5\,PN gravitational–wave radiation reaction \citep{Peters1964,Kidder1995}; 1\,PN and spin--orbit (1.5\,PN) effects are available via the \rev{\texttt{gr\_full}} and \rev{\texttt{lense\_thirring}} modules. Each operator is designed to be unit–agnostic and exposes minimal, physically interpretable parameters that can be calibrated or varied systematically.


Our goal is to make these processes available, documented, and reproducible within a single $N$-body environment so that studies of close-binary evolution can move seamlessly between isolated binaries and dynamically rich contexts. The remainder of this paper details the physical models and  numerical implementation, and example applications.

\section{Effects implementation}
\label{sec:AlgorithmicFlow}
Below we present the physical and numerical implementations of the new binary evolution modules.
We note that multiple examples and comparisons against other codes are provided in the Appendix.

\subsection{Simplified Stellar Evolution}
\label{sec:sse}
Some of the effects implemented below depend on having self-consistent stellar
radii and luminosities as functions of mass. To supply these properties, the
optional \texttt{stellar\_evolution\_sse} operator (Table 1) updates each star's radius
$R$ and luminosity $L$ from analytic, mass-dependent prescriptions \citep{Hurley2000}. The update
is purely algebraic: the timestep is ignored and the relations are evaluated
using the particle's current mass each call. Virtual particles are skipped,
and stellar masses remain unchanged.

The \rev{dimensionless stellar mass} $m\equiv M/M_\odot$ is formed using the operator-level parameter
\texttt{sse\_Msun} (default 1) to convert code masses into solar masses.
\texttt{sse\_Rsun} and \texttt{sse\_Lsun} analogously define the solar radius
and luminosity in code units. The particle's radius
\texttt{sim.particles[index].r} field is set to $R$, and the luminosity $L$ is
written to the particle attribute \texttt{sse\_L}. This stored luminosity can
be used by other modules such as the wind operators.

\paragraph{Basic power-law fallback.}
{ 
If a particle provides any of the basic
power-law parameters \texttt{sse\_R\_coeff}, \texttt{sse\_R\_exp},
\texttt{sse\_L\_coeff}, or \texttt{sse\_L\_exp}, then the operator uses the
power-law mapping
\[
R = R_{\rm coeff}\,R_\odot\,m^{R_{\rm exp}},\qquad
L = L_{\rm coeff}\,L_\odot\,m^{L_{\rm exp}},
\]
with defaults $R_{\rm coeff}=1$, $R_{\rm exp}=0.8$, $L_{\rm coeff}=1$,
$L_{\rm exp}=3.5$ if any of the corresponding particle-level parameters are unspecified.

\paragraph{Object-class presets via \texttt{sse\_type}.}
If no basic power-law parameters are present, the operator uses the
per-particle integer \texttt{sse\_type} to select one of five analytic
structure presets. Two optional per-particle multipliers,
\texttt{sse\_R\_mult} and \texttt{sse\_L\_mult} (defaults 1), rescale the
preset (and fallback) values as
$R\leftarrow R\times\texttt{sse\_R\_mult}$ and
$L\leftarrow L\times\texttt{sse\_L\_mult}$. The presets are:

\begin{itemize}[nosep,leftmargin=1.8em]
\item {\texttt{sse\_type}=1 (H-rich main sequence / ZAMS-like).}
We use continuous piecewise scalings for $R(m)$ and $L(m)$:
\[
\frac{R}{R_\odot}=
\begin{cases}
m^{0.80}, & m\le 1,\\
m^{0.57}, & 1<m\le 10,\\
10^{0.57}\,\left(\dfrac{m}{10}\right)^{0.50}, & m>10,
\end{cases}
\qquad
\frac{L}{L_\odot}=
\begin{cases}
0.23\,m^{2.30}, & m<0.43,\\
m^{4.00}, & 0.43\le m<2,\\
\left(\dfrac{16}{2^{3.5}}\right)\,m^{3.50}, & m\ge 2.
\end{cases}
\]

\item {\texttt{sse\_type}=2 (H-rich giant; RGB/AGB lumped).}
We adopt a large-radius, weak-mass scaling with an $L\propto R^2$ closure
corresponding to a representative cool effective temperature:
\[
\frac{R}{R_\odot}=100\,m^{-0.30},\qquad
\frac{L}{L_\odot}=0.23\left(\frac{R}{R_\odot}\right)^{2}.
\]
The multipliers \texttt{sse\_R\_mult} and \texttt{sse\_L\_mult} allow users to
shift the fiducial giant scale without supplying full evolutionary tracks.

\item {\texttt{sse\_type}=3 (stripped helium star; He-MS / WR-like).}
We use compact, luminous power-law scalings:
\[
\frac{R}{R_\odot}=0.20\,m^{0.60},\qquad
\frac{L}{L_\odot}=50\,m^{2.50}.
\]

\item {\texttt{sse\_type}=4 (white dwarf).}
The radius follows a Chandrasekhar-mass normalised mass--radius relation,
\[
\frac{R}{R_\odot}=R_{\rm wd,coeff}\,
\left[\left(\frac{m}{M_{\rm Ch}}\right)^{-2/3}
      -\left(\frac{m}{M_{\rm Ch}}\right)^{\;\;2/3}\right]^{1/2},
\]
with $M_{\rm Ch}=\texttt{sse\_Mch}$ (default 1.44) and
$R_{\rm wd,coeff}=\texttt{sse\_Rwd\_coeff}$ (default 0.0112). The luminosity is
set to $L/L_\odot=\texttt{sse\_Lwd}$ (default $10^{-3}$) unless rescaled by
\texttt{sse\_L\_mult}.

\item {\texttt{sse\_type}=5 (compact remnant; NS/BH).}
The operator assigns a constant neutron-star radius if $m\le
\texttt{sse\_Mns\_max}$ (default 3), otherwise a Schwarzschild-radius scaling
for a black hole:
\[
\frac{R}{R_\odot}=
\begin{cases}
\texttt{sse\_Rns\_factor}, & m\le \texttt{sse\_Mns\_max},\\
\texttt{sse\_Rbh\_factor}\,m, & m> \texttt{sse\_Mns\_max},
\end{cases}
\qquad
\frac{L}{L_\odot}=0.
\]
The defaults are \texttt{sse\_Rns\_factor}$=1.724\times10^{-5}$ and
\texttt{sse\_Rbh\_factor}$=4.246\times10^{-6}$.
\end{itemize}

If \texttt{sse\_type} is absent or equals 0 and no basic power-law parameters
are present, the operator falls back to the default power laws
$R/R_\odot=m^{0.8}$ and $L/L_\odot=m^{3.5}$.
}

\begin{table}[h]
\centering\footnotesize
\caption{Stellar-structure parameters (\texttt{stellar\_evolution\_sse})}
\label{tab:sse}
\begin{tabular}{@{}llll@{}}
\toprule
Name (scope) & Unit & Default & Purpose \\
\midrule
\texttt{sse\_Msun}   (op) & mass      & 1   & Solar mass in code units ($M_\odot$)\\
\texttt{sse\_Rsun}   (op) & length    & 1   & Solar radius in code units ($R_\odot$)\\
\texttt{sse\_Lsun}   (op) & luminosity& 1   & Solar luminosity in code units ($L_\odot$)\\[0.2em]
\texttt{sse\_type}   (part) & —       & 0   & Object class selector (0--5; see text)\\
\texttt{sse\_R\_mult} (part) & —       & 1   & Radius multiplier for presets\\
\texttt{sse\_L\_mult} (part) & —       & 1   & Luminosity multiplier for presets\\[0.2em]
\texttt{sse\_R\_coeff} (part) & —      & 1   & {basic} radius prefactor (overrides \texttt{sse\_type})\\
\texttt{sse\_R\_exp}   (part) & —      & 0.8 & {basic} radius mass exponent\\
\texttt{sse\_L\_coeff} (part) & —      & 1   & {basic} luminosity prefactor\\
\texttt{sse\_L\_exp}   (part) & —      & 3.5 & {basic} luminosity mass exponent\\[0.2em]
\texttt{sse\_Mch}        (op) & —      & 1.44 & Chandrasekhar mass in $M_\odot$ (WD preset)\\
\texttt{sse\_Rwd\_coeff} (op) & —      & 0.0112 & WD radius coefficient in $R_\odot$\\
\texttt{sse\_Lwd}        (op) & —      & $10^{-3}$ & WD luminosity in $L_\odot$\\
\texttt{sse\_Rns\_factor} (op) & —     & $1.724\times10^{-5}$ & NS radius in $R_\odot$\\
\texttt{sse\_Rbh\_factor} (op) & —     & $4.246\times10^{-6}$ & Schwarzschild radius per $M_\odot$ in $R_\odot$\\
\texttt{sse\_Mns\_max}    (op) & —     & 3.0 & NS/BH boundary mass in $M_\odot$ (compact preset)\\
\texttt{sse\_L} (part, out) & luminosity & — & Stored luminosity used by other operators\\
\bottomrule
\end{tabular}
\end{table}

\subsection{Reimers Stellar Wind Mass Loss}
\label{sec:swml}
In the Reimers prescription \citep{Reimers1975}, isotropic winds remove mass from single cool, low- to intermediate-mass giants. For a star of mass $M$, luminosity $L$, and radius
$R$, the mass-loss rate is
\[
\dot M = -4\times10^{-13}\,\eta\,\frac{L}{L_\odot}\frac{R}{R_\odot}\frac{M_\odot}{M}
\;M_\odot\,\mathrm{yr}^{-1},
\]
where the dimensionless efficiency $\eta$ and luminosity $L$ (\rev{\texttt{sse\_L}} defined above) are specified per particle, while $R$ is taken from the particle's radius field. \rev{Stars lacking either \texttt{swml\_eta} or \texttt{sse\_L} are skipped.}
Mass is removed isotropically with no linear-momentum
recoil; virtual particles are ignored and after mass loss the system is
recentred on the centre of mass.  A safety limiter caps the fractional
mass change to \texttt{swml\_max\_dlnM} (default $0.1$) per call.  The
prefactor, solar reference values, and the year length can be adjusted via
operator parameters (Table~\ref{tab:swml}).  By default the operator
suppresses winds when a particle is flagged as inside a common envelope or
undergoing Roche–lobe overflow; the switches
\texttt{swml\_disable\_in\_CE} and \texttt{swml\_disable\_in\_RLOF}
control this behaviour by reading the per-particle flags
\texttt{inside\_CE} and \texttt{rlof\_active}.

\begin{table}[h]
\centering\footnotesize
\caption{Stellar wind mass-loss parameters}
\label{tab:swml}
\begin{tabular}{@{}llll@{}}
\toprule
Name (scope) & Unit & Default & Purpose \\
\midrule
\texttt{swml\_eta} (part) & — & — & Wind efficiency $\eta$\\
\texttt{swml\_const} (op) & $M_\odot$/yr & $4\times10^{-13}$ & Reimers prefactor\\
\texttt{swml\_Msun}  (op) & mass & 1 & Solar mass in code units\\
\texttt{swml\_Rsun}  (op) & length & 1 & Solar radius in code units\\
\texttt{swml\_Lsun}  (op) & luminosity & 1 & Solar luminosity in code units\\
\texttt{swml\_year}  (op) & time & 1 & Length of Julian year in code units\\
\texttt{swml\_max\_dlnM} (op) & — & 0.1 & Max $|\Delta M|/M$ per call\\
\texttt{swml\_disable\_in\_CE} (op) & bool & 1 & Skip if \texttt{inside\_CE}$>0$\\
\texttt{swml\_disable\_in\_RLOF} (op) & bool & 1 & Skip if \texttt{rlof\_active}$>0$\\
\bottomrule
\end{tabular}
\end{table}

\subsection{Parker-type thermal wind}
\label{sec:tdw}

In cool, low-gravity stars, thermal pressure can launch isotropic winds that carry away mass \citep{parker}. For a star
of mass $M$, luminosity $L$, and radius $R$, we adopt a Parker-like scaling
\[
\dot M = -C_{\rm th}\,\eta\,\left(\frac{R}{R_\odot}\right)^{\alpha_R}
                   \left(\frac{L}{L_\odot}\right)^{\alpha_L}
                   \left(\frac{M_\odot}{M}\right)^{\alpha_M}
\;M_\odot\,\mathrm{yr}^{-1},
\]
where $\eta$ is a dimensionless heating efficiency and the exponents have
defaults $(\alpha_R,\alpha_L,\alpha_M)=(2,\tfrac{3}{2},1)$.  The luminosity
$L$ is read from the particle attribute \texttt{sse\_L}, which must be
populated either by the simplified stellar‑evolution operator or manually by
the user, while $R$ and $M$ are taken from the particle's $r$ and $m$ fields.
\rev{Stars lacking either \texttt{tdw\_eta} or \texttt{sse\_L} are skipped.} Mass is removed isotropically
with no linear-momentum recoil; virtual particles are ignored, and after mass
loss the system is recentred on the centre of mass.  The prefactor, solar
reference values, exponents, maximum fractional mass change, and the year
length can be adjusted via operator parameters (Table~\ref{tab:tdw}). The
operator is unit-agnostic if these scaling constants are specified
consistently.
By default the operator suspends winds for stars marked as being inside a
common envelope or undergoing active Roche--lobe overflow, as indicated by
the per-particle flags \texttt{inside\_CE} and \texttt{rlof\_active}.  This
behaviour is controlled by \texttt{tdw\_disable\_in\_CE} and
\texttt{tdw\_disable\_in\_RLOF}.

\paragraph{Applicability: temperature and surface gravity.}
{
\rev{The Parker-type prescription is intended to represent thermally driven coronal/chromospheric winds.} For an isothermal Parker wind
with wind temperature $T_{\rm w}$, the sonic radius is
\begin{equation}
r_s=\frac{G M}{2 c_s^2}
=\frac{G M \mu m_p}{2 k_B T_{\rm w}},
\end{equation}
where $\mu$ is the mean molecular weight and $m_p$ is the proton mass.
A transonic solution launched near the stellar surface requires $r_s>R_\star$,
i.e.
\begin{equation}
T_{\rm w}<T_{\max}\equiv \frac{G M \mu m_p}{2 k_B R_\star}
=\frac{\mu m_p}{2 k_B}\,g\,R_\star,
\end{equation}
with $g\equiv GM/R_\star^2$ the surface gravity.
In practice, Parker-like winds correspond to $r_s/R_\star\sim 2$--$10$, or
equivalently $T_{\rm w}\sim (0.1$--$0.5)\,T_{\max}$.
This maps to $T_{\rm w}\sim 0.5$--$3~\mathrm{MK}$ for solar-type dwarfs
($\log g\sim 4$--$5$) and $T_{\rm w}\sim 10^{4}$--$10^{5}~\mathrm{K}$ for giants
($\log g\sim 0$--$3$). The operator is not intended for compact objects (very
high $\log g$) or regimes where winds are primarily line-driven or
dust/radiation-pressure driven.}

\begin{table}[h]
\centering\footnotesize
\caption{Thermally driven wind parameters}
\label{tab:tdw}
\begin{tabular}{@{}llll@{}}
\toprule
Name (scope) & Unit & Default & Purpose \\
\midrule
\texttt{tdw\_eta} (part) & — & — & Wind efficiency $\eta$\\
\texttt{sse\_L}   (part) & luminosity & — & Stellar luminosity $L$\\[0.2em]
\texttt{tdw\_const} (op) & $M_\odot$/yr & $2\times10^{-14}$ & Thermal-wind prefactor $C_{\rm th}$\\
\texttt{tdw\_Msun}  (op) & mass & 1 & Solar mass in code units\\
\texttt{tdw\_Rsun}  (op) & length & 1 & Solar radius in code units\\
\texttt{tdw\_Lsun}  (op) & luminosity & 1 & Reference luminosity $L_\odot$ in code units\\
\texttt{tdw\_year}  (op) & time & 1 & Length of Julian year in code units\\
\texttt{tdw\_alpha\_R} (op) & — & 2 & Exponent $\alpha_R$ on $R/R_\odot$\\
\texttt{tdw\_alpha\_L} (op) & — & $3/2$ & Exponent $\alpha_L$ on $L/L_\odot$\\
\texttt{tdw\_alpha\_M} (op) & — & 1 & Exponent $\alpha_M$ on $M_\odot/M$\\
\texttt{tdw\_max\_dlnM} (op) & — & 0.1 & Max $|\Delta M|/M$ per call\\
\texttt{tdw\_disable\_in\_CE} (op) & bool & 1 & Skip if \texttt{inside\_CE}$>0$\\
\texttt{tdw\_disable\_in\_RLOF} (op) & bool & 1 & Skip if \texttt{rlof\_active}$>0$\\
\bottomrule
\end{tabular}
\end{table}

\subsection{Eddington-Limited Winds}
\label{sec:edw}

Luminosities exceeding the electron-scattering Eddington limit in some massive and highly luminous stellar objects drive isotropic
mass loss \citep[e.g.][]{Eddington1926,Shaviv2001} according to
\[
\dot M = -C_{\rm edd}\,\max\!\left(0,\frac{L}{L_{\rm Edd}}-1\right)
\;M_\odot\,\mathrm{yr}^{-1},
\]
where $L_{\rm Edd} = L_{\rm Edd,coeff}\,(M/M_\odot)L_\odot$. The luminosity $L$
is read from the particle attribute \texttt{sse\_L}, which must be supplied
by the user or the simplified stellar-evolution operator. Mass is removed
isotropically with no linear-momentum recoil; virtual particles are ignored,
and after mass loss the system is recentred on the centre of mass. A limiter
caps the fractional mass change at \texttt{edw\_max\_dlnM} per call. Operator
parameters controlling the scaling constants are listed in
Table~\ref{tab:edw}.
By default the operator skips stars flagged as being inside a common envelope
or undergoing Roche--lobe overflow, controlled by
\texttt{edw\_disable\_in\_CE} and \texttt{edw\_disable\_in\_RLOF}, which read
\texttt{inside\_CE} and \texttt{rlof\_active}.

\begin{table}[h]
\centering\footnotesize
\caption{Super-Eddington wind parameters}
\label{tab:edw}
\begin{tabular}{@{}llll@{}}
\toprule
Name (scope) & Unit & Default & Purpose \\
\midrule
\texttt{sse\_L} (part) & luminosity & — & Stellar luminosity $L$\\
\texttt{edw\_const} (op) & $M_\odot$/yr & $1\times10^{-6}$ & Mass-loss prefactor $C_{\rm edd}$\\
\texttt{edw\_Msun}  (op) & mass & 1 & Solar mass in code units\\
\texttt{edw\_Lsun}  (op) & luminosity & 1 & Solar luminosity in code units\\
\texttt{edw\_year}  (op) & time & 1 & Length of Julian year in code units\\
\texttt{edw\_Ledd\_coeff} (op) & $L_\odot/M_\odot$ & $3.2\times10^{4}$ & $L_{\rm Edd}$ per unit mass\\
\texttt{edw\_max\_dlnM} (op) & — & 0.1 & Max $|\Delta M|/M$ per call\\
\texttt{edw\_disable\_in\_CE} (op) & bool & 1 & Skip if \texttt{inside\_CE}$>0$\\
\texttt{edw\_disable\_in\_RLOF} (op) & bool & 1 & Skip if \texttt{rlof\_active}$>0$\\
\bottomrule
\end{tabular}
\end{table}
\subsection{Magnetic Braking}
\label{sec:mb}

Low‑mass stars with convective envelopes lose spin angular momentum through
magnetised winds \citep[e.g.][]{Skumanich1972,Rappaport1983}. In tidally coupled binaries (particularly when a component is close to synchronous rotation), tidal torques replenish the braked spin by drawing from the orbital reservoir, leading to a net loss of orbital angular momentum and a secular decrease of the semi‑major axis and orbital period, whereas in wide or weakly coupled systems the orbital response is negligible on comparable timescales.

We implement the Verbunt--Zwaan / Kawaler torque law
\citep{Verbunt1981,Kawaler1988}, applying a braking torque antiparallel to the
spin vector (parameters in Table 5).

\paragraph{Torque law.}
For a star of mass $M$, radius $R$, and angular velocity
$\omega = \lVert\boldsymbol{\Omega}\rVert$, the spin‑down torque is
\begin{equation}
\label{eq:mb_torque}
\frac{dJ}{dt}
= -K\,\Big(\frac{R}{R_\odot}\Big)^{1/2}\Big(\frac{M}{M_\odot}\Big)^{-1/2}
\begin{cases}
\omega^3, & \omega \le \omega_{\rm sat},\\[3pt]
\omega\,\omega_{\rm sat}^{2}, & \omega > \omega_{\rm sat},
\end{cases}
\end{equation}
where $K$ is a normalisation constant (operator parameter \texttt{mb\_K},
default $2.7\times10^{47}$ in cgs). The torque acts colinearly with
$\boldsymbol{\Omega}$, so the spin direction is unchanged by this operator.

\paragraph{Units and scaling.}
Users specify \texttt{mb\_K} in cgs. Internally we convert to code units using
the operator parameters \texttt{mb\_Msun}, \texttt{mb\_Rsun}, and
\texttt{mb\_year}, which define $M_\odot$, $R_\odot$, and the Julian year in
{code} units. Defining the cgs-per-code base units
$M_{\rm unit}=\mathrm{g}/\mathrm{(code\;mass)}$,
$L_{\rm unit}=\mathrm{cm}/\mathrm{(code\;length)}$,
$T_{\rm unit}=\mathrm{s}/\mathrm{(code\;time)}$, and noting that
$[K]=M\,L^{2}\,T$, we obtain
\begin{equation}
K_{\rm code}=\frac{K_{\rm cgs}}{M_{\rm unit}\,L_{\rm unit}^{2}\,T_{\rm unit}},
\end{equation}
and then evaluate Eq.~\eqref{eq:mb_torque} with the explicit
dimensionless factors $(R/R_\odot)^{1/2}(M/M_\odot)^{-1/2}$.
Users do not need to manually rescale \texttt{mb\_K}.

\paragraph{Saturation.}
If a particle provides a convective turnover time \texttt{mb\_tau\_conv},
the code sets the critical angular velocity via the Rossby number as
\begin{equation}
\omega_{\rm sat} = \frac{2\pi}{\texttt{mb\_Rossby\_sat}\times \texttt{mb\_tau\_conv}},
\end{equation}
with \texttt{mb\_Rossby\_sat} (operator level; default 0.1).
A particle-supplied \texttt{mb\_omega\_sat} overrides this value.
For $\omega>\omega_{\rm sat}$ the torque scales linearly with $\omega$,
ensuring a continuous transition at the threshold.

\paragraph{Time integration (closed form).}
Because the torque is colinear with $\boldsymbol{\Omega}$,
the magnitude $\omega$ obeys a scalar ODE with a closed-form solution. Define
\begin{equation}
C \;=\; \frac{K_{\rm code}}{I}\,
\Big(\frac{R}{R_\odot}\Big)^{1/2}\Big(\frac{M}{M_\odot}\Big)^{-1/2}.
\end{equation}
Then
\begin{align}
\text{unsaturated:}\quad
&\frac{d\omega}{dt} = -C\,\omega^3
\;\Rightarrow\;
\omega(t+\Delta t) = \frac{\omega(t)}{\sqrt{1+2C\,\omega(t)^2\,\Delta t}},
\\[4pt]
\text{saturated:}\quad
&\frac{d\omega}{dt} = -C\,\omega_{\rm sat}^2\,\omega
\;\Rightarrow\;
\omega(t+\Delta t) = \omega(t)\,\exp\!\big[-C\,\omega_{\rm sat}^2\,\Delta t\big].
\end{align}
If a step starts in the saturated regime and crosses to the unsaturated regime,
the update is applied piecewise: exponential decay to
$\omega_{\rm sat}$ followed by the unsaturated formula for the remainder of the
step. The spin vector is finally rescaled by
$\boldsymbol{\Omega}\leftarrow \boldsymbol{\Omega}\,\rev{\bigl[\omega(t+\Delta t)/\omega(t)\bigr]}$.
This scheme is positivity‑preserving and avoids step‑size instabilities.

\paragraph{Activation and guards.}
Braking is applied only if a particle sets \texttt{mb\_on}$=1$ and
\texttt{mb\_convective}$=1$. Missing parameters, non‑positive or non‑finite
$I$, $M$, or $R$ bypass the update. A per‑particle radius override
\texttt{mb\_R} (in code length units) is supported; otherwise the particle
radius \texttt{p->r} is used. The operator does nothing for vanishing spin
vectors.

\begin{table}[h]
\centering\footnotesize
\caption{Magnetic braking parameters}
\label{tab:mb}
\begin{tabular}{@{}llll@{}}
\toprule
Name (scope) & Unit & Default & Purpose \\
\midrule
\texttt{mb\_K} (op) & cgs & $2.7\times10^{47}$ & Braking constant $K$\\
\texttt{mb\_Msun} (op) & mass & 1 & Solar mass in code units\\
\texttt{mb\_Rsun} (op) & length & 1 & Solar radius in code units\\
\texttt{mb\_year} (op) & time & 1 & Julian year in code units\\
\texttt{mb\_Rossby\_sat} (op) & — & 0.1 & Critical Rossby number\\
\texttt{mb\_on} (part) & bool & 0 & Enable magnetic braking\\
\texttt{mb\_convective} (part) & bool & 0 & Convective‑envelope flag\\
\texttt{mb\_omega\_sat} (part) & 1/t & $\infty$ & Saturation angular velocity\\
\texttt{mb\_tau\_conv} (part) & time & — & Convective turnover time\\
\texttt{mb\_R} (part) & length & — & Radius override (else particle radius)\\
\texttt{I} (part) & mass\,length$^2$ & — & Moment of inertia\\
\texttt{Omega} (part, vec) & 1/t & — & Spin angular‑velocity vector (\texttt{reb\_vec3d})\\
\bottomrule
\end{tabular}
\end{table}

\paragraph{Sanity checks.}
With \texttt{mb\_Msun} = \texttt{mb\_Rsun} = \texttt{mb\_year} = 1, $M=R=1$,
and $\omega$ in rad/year, the unsaturated torque scales as $\omega^3$ with the
expected normalisation from the literature. The update is continuous at
$\omega=\omega_{\rm sat}$ by construction.

\subsection{Post-Newtonian (PN) corrections}
\label{sec:pn}

Compact binaries experience relativistic corrections that couple the spins and
radiate orbital energy \citep[e.g.][]{Blanchet2014}. The \texttt{post\_newtonian} effect (Table 6) implements the
harmonic-coordinate point-mass equations of motion from \citet{Kidder1995} for
each massive pair:
\begin{itemize}[nosep,leftmargin=1.8em]
  \item $2$\,PN conservative point-mass (PM) and spin--spin (SS) corrections;
  \item $2.5$\,PN gravitational-wave radiation reaction (RR).
\end{itemize}

\paragraph{Variables and masses.}
For two bodies with masses $m_i$, $m_j$, separation vector $\mathbf{r}$,
relative velocity $\mathbf{v}$, define
\[
  m = m_i + m_j,\quad
  \mu = \frac{m_i m_j}{m},\quad
  \eta = \frac{\mu}{m},\quad
  \mathbf{n} = \frac{\mathbf{r}}{r},\quad
  \dot r = \mathbf{v}\cdot\mathbf{n}.
\]
Spins $S_1,S_2$ are {physical} angular momenta (units mass\,length$^2$/time).
If users prefer dimensionless spins $\chi_i$, convert via
$S_i = \chi_i\,G m_i^2 / c$ in the simulation's unit system.

\paragraph{2\,PN conservative.}
The point-mass part is written as
\[
\mathbf{a}_{2\mathrm{PN}}^{\rm pm}
= -\frac{G\,m}{c^4 r^2}
   \Bigl[A_{2}\,\mathbf{n} + B_{v,2}\,\mathbf{v}\Bigr],
\]
with $A_2$ and $B_{v,2}$ built from $v^2$, $\dot r^2$, and $Gm/r$;
the implementation includes the \(-\tfrac12\,\eta(13-4\eta)\,(Gm/r)\,v^2\) term
so that $A_2$ matches \citet{Kidder1995}. The spin--spin coupling is
\[
  \mathbf{a}_{2\mathrm{PN}}^{\rm ss}
  = -\frac{3\,G}{\mu\,c^2\,r^4}
    \left[
      \bigl(S_1\!\cdot\!S_2 - 5(\mathbf{n}\!\cdot\!S_1)(\mathbf{n}\!\cdot\!S_2)\bigr)\mathbf{n}
      + (\mathbf{n}\!\cdot\!S_2)\,S_1 + (\mathbf{n}\!\cdot\!S_1)\,S_2
    \right],
\]
which uses physical spins and the reduced mass $\mu$ in the overall prefactor.
Hence
\[
  \mathbf{a}_{2\mathrm{PN}}=\mathbf{a}_{2\mathrm{PN}}^{\rm pm}+\mathbf{a}_{2\mathrm{PN}}^{\rm ss}.
\]

\paragraph{2.5\,PN radiation reaction.}
\[
\mathbf{a}_{2.5\mathrm{PN}}
= \frac{8\,G^2 m^2 \eta}{5\,c^5 r^3}
\left[
  \dot r\left(18v^2+\tfrac{2}{3}\tfrac{Gm}{r}-25\dot r^2\right)\mathbf{n}
 -\left(6v^2-2\tfrac{Gm}{r}-15\dot r^2\right)\mathbf{v}
\right].
\]

\paragraph{Units and parameters.}
The effect parameter \texttt{c} is the speed of light in {code units}.
The toggles \rev{\texttt{pn\_2PN}} and \rev{\texttt{pn\_25PN}}
(default 1) enable the corresponding sectors. Particle parameter
\texttt{pn\_spin} is a \texttt{reb\_vec3d} storing the physical spin vector.
\begin{table}[h]
\centering\footnotesize
\caption{Post-Newtonian parameters}
\label{tab:pn}
\begin{tabular}{@{}lll p{6.5cm}@{}}
\toprule
Name (scope) & Unit & Default & Purpose \\
\midrule
\texttt{c} (force) & speed & --- & Speed of light (required) \\
\texttt{pn\_2PN} (force) & bool & 1 & Enable 2\,PN point-mass + spin--spin \\
\texttt{pn\_25PN} (force) & bool & 1 & Enable 2.5\,PN radiation reaction \\
\texttt{pn\_merge\_dist} (force) & length & 0 &
Pre-pass merger distance: if $>0$, merge any pair with $r \le \texttt{pn\_merge\_dist}$ before PN; if $=0$, merge only for exact coincidence. \\
\texttt{pn\_spin} (part, vec) & ang. mom. & --- & Physical spin vector $S$ \\
\bottomrule
\end{tabular}
\end{table}

\paragraph{Momentum conservation and scope.}
Accelerations are built for the relative coordinate and split between bodies
in proportion to their masses, preserving total linear momentum.
Spin precession equations are not included; if spins are misaligned, the
orbital dynamics include spin--spin forces with fixed spins.

\paragraph{Effect on binary orbits.}
The 2.5\,PN term removes orbital energy and angular momentum, driving a secular
decrease of the semi-major axis and (typically) eccentricity toward merger.
When spins are present, the spin–spin terms cause relativistic
precession and can modulate the instantaneous orbital plane and pericentre
orientation, altering waveform phase evolution even when the mean orbital
elements change slowly.

\subsection{Roche–Lobe Overflow and Common–Envelope Operator}
\label{sec:RLOF}

This effect models mass transfer in compact binaries via Roche–lobe overflow
(RLOF) together with optional common–envelope (CE) drag  (parameters in Table 7). The implementation
targets robust long integrations: internal sub–stepping constrains
$|\Delta M|/M$ and $|\Delta r|/r$, merge guards prevent divergences, and
diagnostics expose the accumulated energy change from non–Hamiltonian updates.
After each sub–step, the simulation is moved to the updated centre of mass so
that the orbital elements reflect the redistributed mass and momentum.

\paragraph{Roche geometry and mass–loss law.}
The donor’s volume–equivalent Roche radius is evaluated with the Eggleton
formula \citep{Eggleton1983},
\begin{equation}
R_{\rm L} = r\,\frac{0.49\,q^{2/3}}{0.60\,q^{2/3} + \ln(1+q^{1/3})},
\qquad q \equiv \frac{M_{\rm d}}{M_{\rm a}},
\end{equation}
where $r$ is the instantaneous donor--accretor separation, and the instantaneous
mass–loss rate follows the Ritter prescription \citep{Ritter1988},
\begin{equation}
\dot M_{\rm d} = -\dot M_0\,
\exp\!\left[\frac{R_\ast - R_{\rm L}}{H_P}\right],
\end{equation}
with the exponent clamped in magnitude ($\bigl|(R_\ast-R_{\rm L})/H_P\bigr|\le 80$) to avoid numerical overflow. The donor
particle provides $H_P$ and $\dot M_0$ as \texttt{rlmt\_Hp} and
\texttt{rlmt\_mdot0}.

\paragraph{Conservative and systemic mass channels.}
Over a sub–step of size $\Delta t$, the donor loses $m_{\rm loss}=-\dot M_{\rm d}\,\Delta t>0$,
which is split into accreted and wind parts:
\[
m_{\rm acc} = (1-f_{\rm loss})\,m_{\rm loss},\qquad
m_{\rm wind} = f_{\rm loss}\,m_{\rm loss},\quad f_{\rm loss}\in[0,1].
\]
The accretor’s {net} mass gain over the sub-step is $m_{\rm acc}$; the wind mass
$m_{\rm wind}$ leaves the system.
For \texttt{jloss\_mode}$=1$ (isotropic re-emission), the code implements this as
a two-stage update: the full $m_{\rm loss}$ is first transferred to the accretor,
and $m_{\rm wind}$ is then expelled from the accretor isotropically; the net
retained mass is still $m_{\rm acc}$.

\paragraph{$j$--loss (wind angular momentum).}
The specific angular momentum carried by the wind is prescribed by a mode
parameter:
\begin{align*}
\text{mode 0:}~& \mathbf v_{\rm loss}=\mathbf v_{\rm d}
\ \ \ (\text{wind leaves from the donor; implicit AM loss via donor mass decrement}), \\
\text{mode 1:}~&
\text{isotropic re-emission from the accretor: }
m_{\rm loss}\ \text{is transferred, then } \\
& \qquad m_{\rm wind}\ \text{is ejected isotropically from the accretor} \\
& \qquad (\text{no recoil in the accretor frame; orbital AM of the expelled mass is removed} \\
& \qquad \text{implicitly by the accretor mass decrement}),\\
\text{mode 2:}~& \mathbf v_{\rm loss}=\mathbf v_{\rm CM},
\ \ \ (\text{wind removes centre-of-mass momentum } m_{\rm wind}\mathbf v_{\rm CM}; \\
& \hspace{2.4em}\text{the donor’s mass decrement still subtracts } m_{\rm wind}(\mathbf r_{\rm d}-\mathbf R_{\rm CM})\times\mathbf v_{\rm d}) ,\\
\text{mode 3 (target-$j$):}~& |\mathbf v_{\rm loss}-\mathbf v_{\rm emit}| 
= f_j\,j_{\rm orb}/r \quad\text{along }\hat e_\perp\!\perp\hat{\mathbf r},\\
& j_{\rm orb}\equiv J/M=\mu\,|\mathbf r\times\mathbf v_{\rm rel}|/(M_{\rm d}+M_{\rm a}),\quad
\mu=\tfrac{M_{\rm d}M_{\rm a}}{M_{\rm d}+M_{\rm a}},\\
& \Delta\mathbf L_{\rm wind}
= m_{\rm wind}(\mathbf r_{\rm emit}-\mathbf R_{\rm CM})\times(\mathbf v_{\rm loss}-\mathbf v_{\rm emit}),\\
& |\Delta\mathbf L_{\rm wind}| 
= m_{\rm wind}f_j j_{\rm orb}\,\frac{|\mathbf r_{\rm emit}-\mathbf R_{\rm CM}|}{r}.
\end{align*}
The torque direction is generally not forced to align with the orbital
angular momentum. If strict alignment or the full 
$m_{\rm wind}f_j j_{\rm orb}$ magnitude is required, interpret $f_j$ as an
effective scale factor $r/|\mathbf r_{\rm emit}-\mathbf R_{\rm CM}|$ and/or
constrain $\hat e_\perp$ to the orbital plane.

\textit{Notes.} For mode 0, no additional wind torque is applied because
$\mathbf v_{\rm loss}=\mathbf v_{\rm emit}$ and the relevant angular momentum is
removed implicitly by the donor mass decrement. For mode 1, the expelled mass is
removed directly from the accretor (isotropic re-emission), so its orbital
angular momentum is removed implicitly by the accretor mass decrement and no
explicit “wind–$\Delta L$” correction is applied. For mode 3, an explicit
wind–$\Delta L$ correction acts through $\mathbf v_{\rm loss}-\mathbf v_{\rm emit}$.
For mode 2, the explicit wind–$\Delta L$ term vanishes by construction when the
emission point is taken at the centre of mass.

\paragraph{Conservative transfer angular momentum.}
In the absence of external torques, conservative mass relocation should not
change the system’s orbital angular momentum. The operator computes the
angular–momentum difference between placing the transferred mass $m_{\rm trans}$
at the donor and at the accretor (both at the donor velocity) and applies a
minimal pure‐torque velocity correction to enforce $\Delta L_{\rm trans}=0$.
Here $m_{\rm trans}=m_{\rm acc}$ for modes 0, 2, and 3, while for isotropic
re-emission (mode 1) the transfer stage uses $m_{\rm trans}=m_{\rm loss}$ and
the subsequent ejection is handled separately at the accretor.

\paragraph{Linear momentum.}
Conservative accretion is treated internally: the transferred mass $m_{\rm trans}$
is added to the accretor with the donor’s instantaneous velocity, while the donor’s
mass is reduced; this leaves the pair’s total momentum unchanged for the transfer
stage.

For modes 0, 2, and 3, the systemic wind mass $m_{\rm wind}$ is removed directly from
the donor, so the donor’s mass decrement already removes $m_{\rm wind}\,\mathbf v_{\rm d}$.
The operator then applies a uniform shift
$-\,m_{\rm wind}(\mathbf v_{\rm loss}-\rev{\mathbf v_{\rm d}})$ to the pair (distributed
as a common velocity increment), so that the total momentum change for these modes is
\[
\Delta \mathbf P = -\,m_{\rm wind}\,\mathbf v_{\rm d}
-\,m_{\rm wind}(\mathbf v_{\rm loss}-\rev{\mathbf v_{\rm d}})
\rev{= -\,m_{\rm wind}\,\mathbf v_{\rm loss}},
\]
i.e.\ the system loses exactly the momentum carried by the escaping wind in the inertial frame.

For mode 1 (isotropic re-emission), the wind is implemented as an accretor mass loss:
after transferring $m_{\rm loss}$ to the accretor, the code removes $m_{\rm wind}$ from the
accretor isotropically (no recoil in the accretor frame). The corresponding momentum
loss is therefore implicit through the accretor mass decrement, with
$\Delta\mathbf P_{\rm wind}=-m_{\rm wind}\,\mathbf v_{\rm a}$ evaluated at the time of
ejection.

After the mass and velocity updates each sub-step the simulation is moved to
the centre of mass to keep the reference frame consistent.

\paragraph{Inter-module flags.}
After each sub-step the operator writes the boolean attributes
\texttt{inside\_CE} and \texttt{rlof\_active} on both donor and accretor.
These indicate whether the accretor lies within the donor's radius and
whether the RLOF channel is actively removing mass. Other effects, such as
stellar winds, inspect these flags (in conjunction with their own disable
switches) to suspend their action during common-envelope phases or while
RLOF is operating.

\subsubsection{Common–Envelope (CE) Drag}
When the accretor is inside the donor ($r<R_\ast$), the code can apply
dynamical friction following Ostriker’s formula \citep{Ostriker1999,Ivanova2013},
\[
\mathbf a_{\rm DF} = -\,\frac{4\pi G^2 M_{\rm a}\rho}{v_{\rm rel}^3}\,
I(\mathcal M)\,\mathbf v_{\rm rel},\qquad\mathcal M=\frac{v_{\rm rel}}{c_s},
\]
with
\[
I(\mathcal M)=
\begin{cases}
\frac{1}{3}\mathcal M^3+\frac{1}{5}\mathcal M^5, & \mathcal M<0.02,\\[0.4em]
\frac{1}{2}\ln\!\frac{1+\mathcal M}{1-\mathcal M}-\mathcal M, & 0.02\le\mathcal M<1,\\[0.4em]
\ln(1/x_{\min})+\tfrac12\ln\!\bigl(1-\mathcal M^{-2}\bigr), \quad \mathcal M\ge1,
\end{cases}
\]
capped by $\ln(1/x_{\min})$ for continuity. A geometric term
$-\pi\rho R_{\rm a}^2 Q_d\,v_{\rm rel}\,\mathbf v_{\rm rel}/M_{\rm a}$ may be
added. The envelope structure comes either from a user–supplied table
$(s,\rho,c_s)$ (\texttt{ce\_profile\_file}) with log–log interpolation or from
a power law $\rho=\rho_0 s^{\alpha_\rho}$, $c_s=c_{s0}s^{\alpha_{c_s}}$.
The velocity kick per sub–step is limited by
$|\Delta\mathbf v|\le\texttt{ce\_kick\_cfl}\,c_s$.
By default the envelope is an {external} sink: no opposite reaction is
applied to the donor; this can be enabled with \texttt{ce\_reaction\_on\_donor}=1.

\subsubsection{Numerics and flow}
Each operator call subdivides the requested interval into at least
\texttt{rlmt\_min\_substeps} sub–steps; the size adapts to satisfy
$|\Delta M|/M\le\texttt{rlmt\_substep\_max\_dm}$ and
$|\Delta r|/r\le\texttt{rlmt\_substep\_max\_dr}$. A merge guard triggers when
$r\le\varepsilon_{\rm merge}$ (user parameter \texttt{merge\_eps} or the
default $0.5\min(R_\ast,R_{\rm a})$ with a small positive floor). On merger the
accretor is removed and the operator detaches to avoid index drift in $N$-body
systems. The operator detaches itself if either component vanishes (non-positive
mass), but it does {not} delete unrelated massless tracer particles.
After mass or drag updates the system is recentered on the centre of mass.
If the optional \texttt{stellar\_evolution\_sse} operator is present, it is
invoked with a zero timestep after RLOF mass updates so that changes in mass
immediately refresh the stellar radii and luminosities.
The accumulated change in the donor--accretor two-body orbital energy over the
call is exported as \texttt{rlmt\_last\_dE} for diagnostics; because the scheme
includes physical sinks (wind, drag) this value is not expected to vanish.

\subsubsection{Dynamical stability and Eddington-limited accretion}
{The \texttt{roche\_lobe\_mass\_transfer} operator does not impose an explicit
analytic stability criterion for RLOF. Instead, the character of the mass transfer (stable,
runaway, or leading to contact) emerges from the time-dependent coupled
evolution of the donor radius, Roche geometry, and orbital response under the
applied mass and momentum updates.}

{Optional Eddington-limited accretion.}
{ 
The operator can optionally enforce a maximum {net} accretion rate onto the
accretor, intended to represent the Eddington limit (or any user-defined cap).
If the accretor supplies \texttt{rlmt\_mdot\_edd} (particle parameter; units
mass/time in the simulation's unit system), then during each internal sub-step
the retained mass is limited to
$m_{\rm acc}\le \texttt{rlmt\_mdot\_edd}\,\Delta t$.
Any excess $\Delta m_{\rm rej}$ above this cap is automatically treated as
non-conservative mass loss and is removed isotropically from the accretor
(isotropic re-emission; no recoil in the accretor frame). This cap acts in
addition to the user-controlled systemic loss fraction
\texttt{rlmt\_loss\_fraction}. If \texttt{rlmt\_mdot\_edd} is absent or
$\le 0$, the cap is disabled and the operator reverts to fully user-controlled
(non-)conservative transfer.
}

\subsubsection{Parameters}
\begin{table}[h]
\centering\footnotesize
\caption{Roche lobe overflow and common envelope evolution parameters.}
\begin{tabular}{@{}lll p{6.5cm}@{}}
\toprule
Name (scope) & Unit & Default & Purpose \\
\midrule
\texttt{rlmt\_donor} (op) & — & — & Donor index (double, cast to int) \\
\texttt{rlmt\_accretor} (op) & — & — & Accretor index (double, cast to int) \\
\texttt{rlmt\_Hp} (part) & length & — & Pressure scale height (donor) \\
\texttt{rlmt\_mdot0} (part) & M/t & — & Reference overflow rate (donor) \\
\texttt{rlmt\_mdot\_edd} (part, accretor) & M/t & 0 &
Maximum allowed net accretion rate (Eddington cap). If $>0$, the operator caps
$m_{\rm acc}\le \texttt{rlmt\_mdot\_edd}\,\Delta t$ and re-emits any excess
isotropically from the accretor. \\
\texttt{rlmt\_loss\_fraction} (op) & — & 0 & Wind fraction $f_{\rm loss}$ \\
\texttt{jloss\_mode} (op) & — & 0 & Wind $j$ prescription (0–3) \\
\texttt{jloss\_factor} (op) & — & 1 & Scale $f_j$ for mode 3 \\
\texttt{rlmt\_skip\_in\_CE} (op) & bool & 1 & Skip RLOF if $r<R_\ast$ \\
\texttt{rlmt\_substep\_max\_dm} (op) & — & $10^{-3}$ & Max $|\Delta M|/M$ per sub–step \\
\texttt{rlmt\_substep\_max\_dr} (op) & — & $5\times10^{-3}$ & Max $|\Delta r|/r$ per sub–step \\
\texttt{rlmt\_min\_substeps} (op) & int & 3 & Minimum sub–steps \\
\texttt{ce\_profile\_file} (op) & path & — & Table $(s,\rho,c_s)$ for CE \\
\texttt{ce\_rho0} (op) & dens & — & CE density normalisation \\
\texttt{ce\_alpha\_rho} (op) & — & 0 & Density slope \\
\texttt{ce\_cs} (op) & speed & — & CE sound-speed normalisation \\
\texttt{ce\_alpha\_cs} (op) & — & 0 & Sound-speed slope \\
\texttt{ce\_xmin} (op) & — & $10^{-4}$ & Coulomb cutoff \\
\texttt{ce\_Qd} (op) & — & 0 & Geometric drag coefficient \\
\texttt{ce\_kick\_cfl} (op) & — & 1 & Velocity–kick limiter \\
\texttt{ce\_reaction\_on\_donor} (op) & bool & 0 & Apply opposite CE kick to donor \\
\texttt{merge\_eps} (op) & length & $0.5\min(R_\ast,R_{\rm a})$ & Merge radius (with floor) \\
\texttt{rlmt\_last\_dE} (op, out) & energy & — & Energy change (diagnostic) \\
\bottomrule
\end{tabular}
\end{table}

\paragraph{Notes.}
All scalar parameters are read as doubles by REBOUNDx and cast to integers for
indices/toggles. If available in the build, a filename in
\texttt{ce\_profile\_file} can be provided to load a CE table; otherwise the
power–law model is used.


\section{Summary \& conclusions}
\label{sec:summary}

\rev{We have presented a suite of binary–evolution modules} for the \textsc{REBOUNDx}
extension to the \textsc{REBOUND} $N$-body integrator that \rev{encapsulate} the
dominant mechanisms driving the orbital and secular evolution of close binaries. The
modules comprise: (i) a Roche–lobe overflow and common–envelope operator that
combines conservative mass transfer with configurable non–conservative
specific–$j$ mass loss and Mach-dependent dynamical friction; (ii) a simplified
stellar–evolution prescription that supplies mass-dependent radii and
luminosities; (iii) three wind channels (Reimers, thermally driven,
and super–Eddington) that remove mass isotropically using those stellar
properties; (iv) a magnetic–braking operator implementing the
Verbunt–Zwaan/Kawaler torque with a saturation-aware closed-form spin update;
and (v) a post-Newtonian module providing 2\,PN point–mass and spin–spin
corrections together with 2.5\,PN gravitational–wave radiation reaction.

Algorithmically, the RLOF/CE operator tracks linear- and angular–momentum
exchanges between the donor, accretor, and escaping mass, enforcing 
linear–momentum conservation for conservative transfer and applying a minimal
torque to maintain angular–momentum consistency. Sub–stepping constraints on
fractional changes in mass and separation, merge guards, and kick limiters for
CE drag stabilise the evolution near contact. Inter-module flags propagate the
onset of RLOF and immersion in a common envelope to the wind operators, while
the simplified stellar evolution is updated in lockstep with mass changes. All
effects are unit-agnostic, relying on user-supplied solar and time scales for
conversion between physical and code units. Together, these ingredients enable
self-consistent, time-resolved investigations of close binaries within
isolated systems or in dynamically rich $N$-body environments.

\section*{Data availability}
 
\rev{The data and code used for this work are available for download from GitHub.}

\section*{Appendix: code tests and examples}
\subsection*{\rev{2.5\,PN} Post-Newtonian radiation reaction: \rev{\textsc{REBOUNDx}} vs \rev{\textsc{phi-GPU}}}
\begin{figure}[h!]
    \centering
    \includegraphics[width=0.8\linewidth]{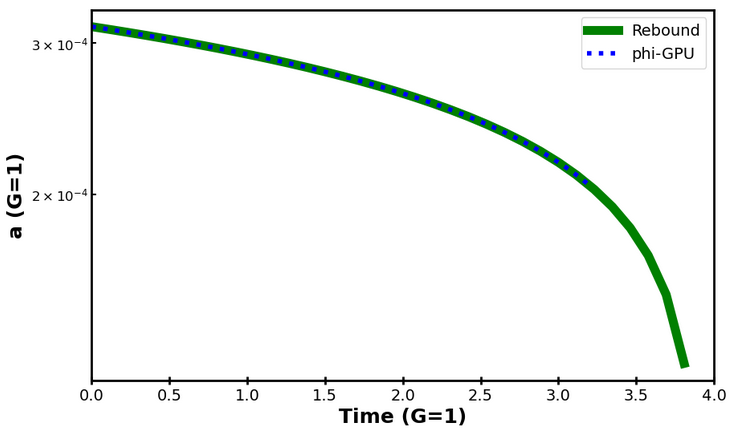}
    \caption{The evolution of the binary \rev{semi-major axis} due to \rev{2.5\,PN} radiation reaction, in \rev{\textsc{REBOUNDx}} (this work) and \rev{\textsc{phi-GPU}}.}
    \label{fig:myplot11}
\end{figure}

In Figure \ref{fig:myplot11} we show an example for the evolution of the binary \rev{semi-major axis} due to the \rev{2.5\,PN} radiation reaction. The system is initialized with $a=0.000313~\mathrm{AU}$, then proceeds to \rev{lose} energy and spiral inward \rev{until} it merges. We moreover show a comparison with the \rev{\textsc{phi-GPU}} code \citep{phi1,phi2}

\begin{lstlisting}[language=Python, caption={\rev{2.5\,PN} Python example (\textsc{REBOUNDx})}]
import rebound
import reboundx

sim = rebound.Simulation()
sim.G=1

sim.add(m=1.00000001E-01,x=   -2.86636483E+00,y= -3.51352660E+00,z=  1.31048588E+00,\
        vx=-8.29922019E+00,vy= -2.77660036E+00,vz= -1.16069792E-01)
sim.add(m=5.99999987E-02,x=   -2.86626798E+00,y= -3.51381408E+00,z=  1.31048970E+00,\
        vx=1.38346340E+01,vy=  4.61131049E+00,vz=  1.91195042E-01)
sim.move_to_com()

rebx = reboundx.Extras(sim)

(*@\textcolor{black}{pn}@*) = rebx.load_force('post_newtonian')
rebx.add_force((*@\textcolor{black}{pn}@*))
(*@\textcolor{black}{pn}@*).params['pn_2PN']=0
(*@\textcolor{black}{pn}@*).params['pn_25PN']=1
(*@\textcolor{black}{pn}@*).params['c']=457.14

sim.integrate(4.2)

\end{lstlisting}

\subsection*{Magnetic braking with tides}
\begin{figure}[h!]
    \centering
    \includegraphics[width=0.8\linewidth]{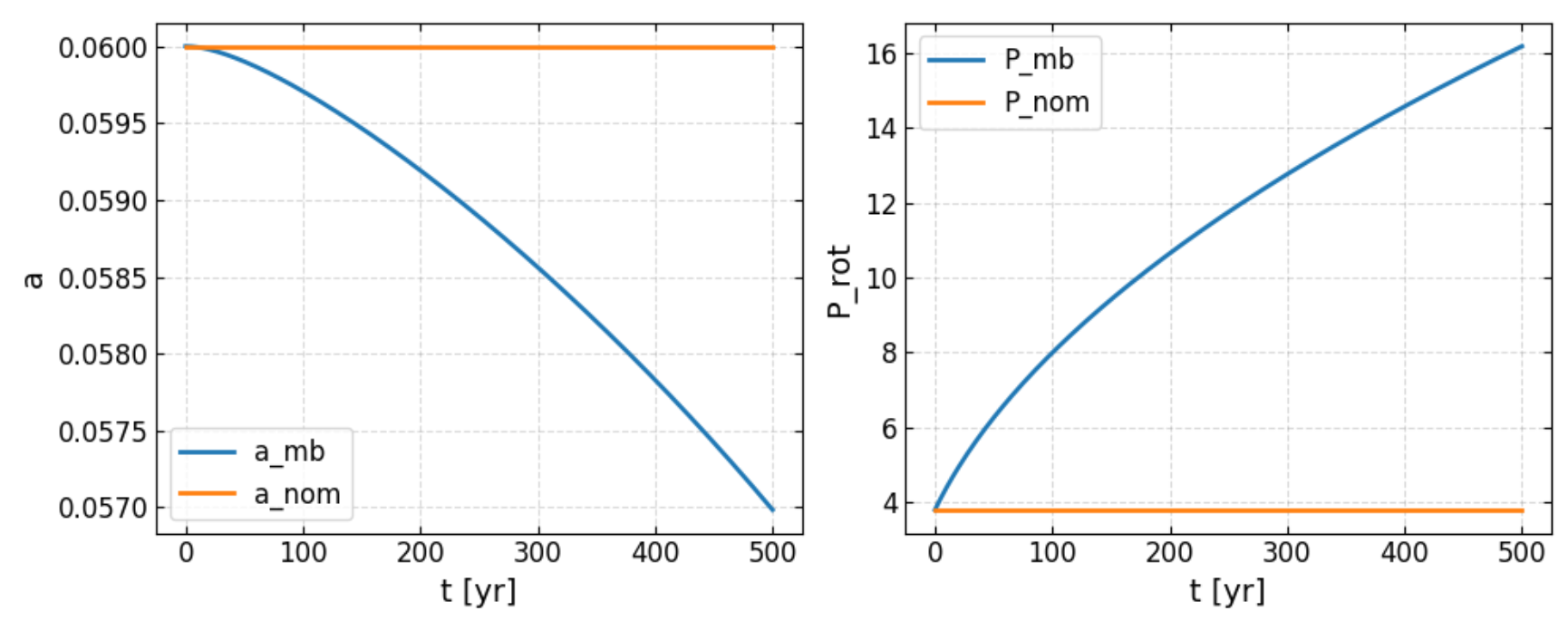}
    \caption{Left panel: evolution of the binary's \rev{semi-major axis} due to a combination of tides and magnetic braking ($_{\rm mb}$) compared to a nominal case without magnetic braking ($_{\rm nom}$). Right panel: evolution of the rotation period of the primary in the presence and absence of magnetic braking.}
    \label{fig:myplot2}
\end{figure}

In Figure \ref{fig:myplot2} we show an example for the evolution of a binary system due to Magnetic braking with tides.

{In this setup the ``no magnetic braking'' run stays essentially unchanged because the orbit starts circular and both stars start exactly synchronous, so tides have almost nothing to damp and the semi-major axis and rotation periods remain nearly constant. When magnetic braking is turned on, it immediately spins the stars down, making them slightly sub-synchronous; tides then act to restore synchronism by transferring angular momentum from the orbit into the stellar spins, but magnetic braking keeps removing that spin angular momentum to an external sink. The combined effect is a secular drain of orbital angular momentum, so the semi-major axis steadily shrinks compared to the no-MB case, while the stellar rotation periods tend to grow (spin down) relative to the no-MB run.}

\begin{lstlisting}[language=Python, caption={Magnetic braking + tides Python example}]
import numpy as np
import rebound
import reboundx

# ==========================================================
# Helper: build one binary simulation
# ==========================================================
def make_binary(with_mb=True):
    sim = rebound.Simulation()
    sim.units = ('AU', 'yr', 'Msun')
    sim.integrator = "ias15"
    sim.ri_ias15.epsilon=0
    rebx = reboundx.Extras(sim)

    # ---------- Binary ICs ----------
    AU_per_Rsun = 1.0/215.032
    m1 = m2 = 1.0
    a0 = 0.06
    e0 = 0.0     # circular orbit
    R1 = R2 = 1.0 * AU_per_Rsun

    # Equal-mass binary in COM frame (for e0=0, this is circular)
    vfac = np.sqrt(sim.G*(m1+m2)/a0) * np.sqrt((1.0 - e0)/(1.0 + e0))
    sim.add(m=m1, r=R1,
            x=-a0*m2/(m1+m2),
            vy=-vfac*m2/(m1+m2))
    sim.add(m=m2, r=R2,
            x=+a0*m1/(m1+m2),
            vy=+vfac*m1/(m1+m2))
    sim.move_to_com()

    # Smaller dt for tighter orbit
    sim.dt = sim.particles[1].P / 20.0

    # ---------- Spins & structure ----------
    k2_moi = 0.7
    I1 = k2_moi*m1*R1**2
    I2 = k2_moi*m2*R2**2

    # Make both stars initially synchronous with the orbit
    P_orb_yr = sim.particles[1].P
    omega_sync = 2.0*np.pi / P_orb_yr  # rad/yr

    for p, I, R in zip(sim.particles[:2], [I1, I2], [R1, R2]):
        p.params["Omega"] = np.array([0.0, 0.0, omega_sync])
        p.params["I"]     = I
        p.r               = R

        # Tidal parameters (used by tides_spin or tctl)
        p.params["k2"]  = 0.2
        p.params["tau"] = 0.01/365.25   # 0.01 days; moderately strong tides

    # ---------- Magnetic braking ----------
    if with_mb:
        mb = rebx.load_operator("magnetic_braking")
        rebx.add_operator(mb)

        mb.params["mb_Msun"]       = 1.0
        mb.params["mb_Rsun"]       = AU_per_Rsun
        mb.params["mb_year"]       = 1.0
        mb.params["mb_Rossby_sat"] = 0.1

        # Stronger than physical, but good for a demo
        mb.params["mb_K"] = 1e55   # cgs; default is ~2.7e47

        for p in sim.particles[:2]:
            p.params["mb_on"]         = 1       # enable MB on this star
            p.params["mb_convective"] = 1
            p.params["mb_tau_conv"]   = 12.5/365.25   # ~12.5 days
    else:
        # No MB operator at all in the "no MB" run
        print("No magnetic_braking operator added for this run")

    # ---------- Tides ----------
    use_spin = True
    try:
        tides = rebx.load_force("tides_spin")
        rebx.add_force(tides)
        use_spin = True
        print("Loaded tides_spin for", "MB" if with_mb else "no-MB", "run")
    except Exception:
        print("tides_spin not available; using tides_constant_time_lag for",
              "MB" if with_mb else "no-MB", "run")
        use_spin = False
        try:
            tides = rebx.load_force("tides_constant_time_lag")
            rebx.add_force(tides)
        except Exception:
            tides = rebx.load_operator("tides_constant_time_lag")
            rebx.add_operator(tides)

        # tctl parameters
        for p in sim.particles[:2]:
            p.params["tctl_k2"]  = 0.2
            p.params["tctl_tau"] = 0.1/365.25
            Om = p.params["Omega"]
            p.params["OmegaMag"] = float(np.linalg.norm(Om))

    return sim, use_spin, R1, R2


# ==========================================================
# Utility functions
# ==========================================================
def elems(sim):
    """Return (a, e) of secondary relative to primary."""
    p1 = sim.particles[1]
    p0 = sim.particles[0]
    o = p1.orbit(primary=p0)
    return o.a, o.e

def prot_days(sim, i):
    """Rotation period of particle i in days."""
    w = np.linalg.norm(sim.particles[i].params["Omega"])
    return (2.0*np.pi/w)*365.25


# ==========================================================
# Build two simulations: with MB and without MB
# ==========================================================
sim_mb,  use_spin_mb,  R1_mb,  R2_mb  = make_binary(with_mb=True)
sim_nom, use_spin_nom, R1_nom, R2_nom = make_binary(with_mb=False)

# ==========================================================
# Integrate both and compare
# ==========================================================
t_end = 500      # years
N     = 200
times = np.linspace(0.0, t_end, N)

for t in times:
    # --- With magnetic braking ---
    sim_mb.integrate(t)

    # --- Without magnetic braking ---
    sim_nom.integrate(t)


\end{lstlisting}

\subsection*{\rev{Magnetic braking} with tides: \rev{\textsc{REBOUNDx}} vs MESA}

\begin{figure}[h!]
    \centering
    \includegraphics[width=0.7\linewidth]{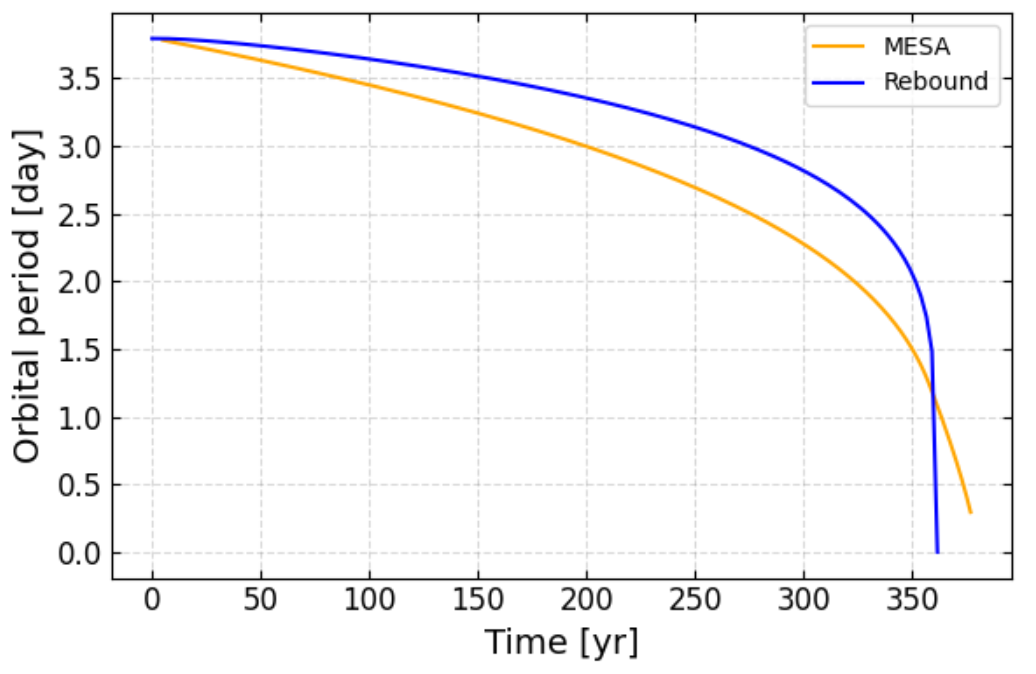}
    \caption{Evolution of the binary's orbital period due to a combination of tides and magnetic braking, in our code and MESA}
    \label{fig:myplot4}
\end{figure}

In Figure \ref{fig:myplot4} we show an example for the evolution of a binary system due to Magnetic braking with tides, compared to MESA \citep{mesa1,mesa2,mesa3,mesa4}.

{To generate the MESA track shown in Figure \ref{fig:myplot4}, we implemented a custom binary ``extras'' module (\texttt{run\_binary\_extras}) that overrides MESA's magnetic-braking angular-momentum loss by wiring a user-defined callback into the binary driver. In \texttt{extras\_binary\_controls} we attach \texttt{other\_jdot\_mb}, which at every step computes a magnetic-braking torque treated as an {orbital} AML term under the synchronous-coupling approximation and assigns it directly to \texttt{b\%jdot\_mb}. The routine \texttt{compute\_mb} evaluates $\omega_{\rm orb}=2\pi/P_{\rm orb}$ from the instantaneous orbital period and defines a saturation threshold $\omega_{\rm sat}$ either from an explicit override or from a Rossby prescription $\omega_{\rm sat}=2\pi/(Ro_{\rm sat}\,\tau_{\rm conv})$, with the normalization $K$, $\tau_{\rm conv}$, and $Ro_{\rm sat}$ passed through \texttt{x\_ctrl} (and sensible defaults adopted if unset). The total $\dot{J}_{\rm MB}$ is then formed by summing the contributions from both stars, scaling as $(R/R_\odot)^{1/2}(M/M_\odot)^{-1/2}$ and switching from an unsaturated $\propto \omega^3$ regime to a saturated $\propto \omega\,\omega_{\rm sat}^2$ regime when $\omega_{\rm orb}>\omega_{\rm sat}$. For transparency and debugging, we additionally record three extra \texttt{binary\_history} columns (\texttt{jdot\_mb\_custom}, \texttt{omega\_orb}, and \texttt{omega\_sat}) to enable a direct diagnostic comparison between the imposed braking law and the resulting binary evolution.}

\begin{lstlisting}[language=Python, caption={\rev{Magnetic braking + tides} Python example (\textsc{REBOUNDx})}]
import numpy as np
import rebound
import reboundx

# --- Parameters ---
K2 = 0.07
K2_MOI = 0.25
TAU_CONV_YR = 12.5 / 365.25
TAU_TIDE_YR = 0.1 / 365.0
AU_PER_RSUN = 1.0 / 215.032


def build_binary_mb() -> rebound.Simulation:
    """Equal-mass circular binary with tides_spin + magnetic braking."""
    sim = rebound.Simulation()
    sim.units = ("AU", "yr", "Msun")
    sim.integrator = "whfast"

    rebx = reboundx.Extras(sim)

    # --- Binary ICs ---
    m1 = m2 = 1.0
    a0 = 0.06
    e0 = 0.0
    R1 = R2 = 1.0 * AU_PER_RSUN

    v = np.sqrt(sim.G * (m1 + m2) / a0) * np.sqrt((1.0 - e0) / (1.0 + e0))
    sim.add(m=m1, r=R1, x=-a0 * m2 / (m1 + m2), vy=-v * m2 / (m1 + m2))
    sim.add(m=m2, r=R2, x=+a0 * m1 / (m1 + m2), vy=+v * m1 / (m1 + m2))
    sim.move_to_com()

    orb0 = sim.particles[1].orbit(primary=sim.particles[0])
    sim.dt = orb0.P / 10.0

    # --- Spins / structure (start synchronous) ---
    omega_sync = 2.0 * np.pi / orb0.P
    I1 = K2_MOI * m1 * R1**2
    I2 = K2_MOI * m2 * R2**2

    for p, I in zip(sim.particles[:2], (I1, I2)):
        p.params["Omega"] = np.array([0.0, 0.0, omega_sync])
        p.params["OmegaMag"] = float(np.linalg.norm(p.params["Omega"]))
        p.params["I"] = I
        p.params["k2"] = K2
        p.params["tau"] = TAU_TIDE_YR
        p.params["tctl_k2"] = K2
        p.params["tctl_tau"] = TAU_TIDE_YR

    # --- Magnetic braking ---
    mb = rebx.load_operator("magnetic_braking")
    rebx.add_operator(mb)
    mb.params["mb_Msun"] = 1.0
    mb.params["mb_Rsun"] = AU_PER_RSUN
    mb.params["mb_year"] = 1.0
    mb.params["mb_Rossby_sat"] = 0.1
    mb.params["mb_K"] = 1e55  # demo-strength

    for p in sim.particles[:2]:
        p.params["mb_on"] = 1
        p.params["mb_convective"] = 1
        p.params["mb_tau_conv"] = TAU_CONV_YR

    # --- Tides ---
    tides = rebx.load_force("tides_spin")
    rebx.add_force(tides)

    return sim


if __name__ == "__main__":
    sim = build_binary_mb()
    t_end_yr = 500.0
    n_steps = 200
    for t in np.linspace(0.0, t_end_yr, n_steps):
        sim.integrate(t)



\end{lstlisting}

\subsection*{Roche lobe overflow}
\begin{figure}[h!]
    \centering
    \includegraphics[width=0.95\linewidth]{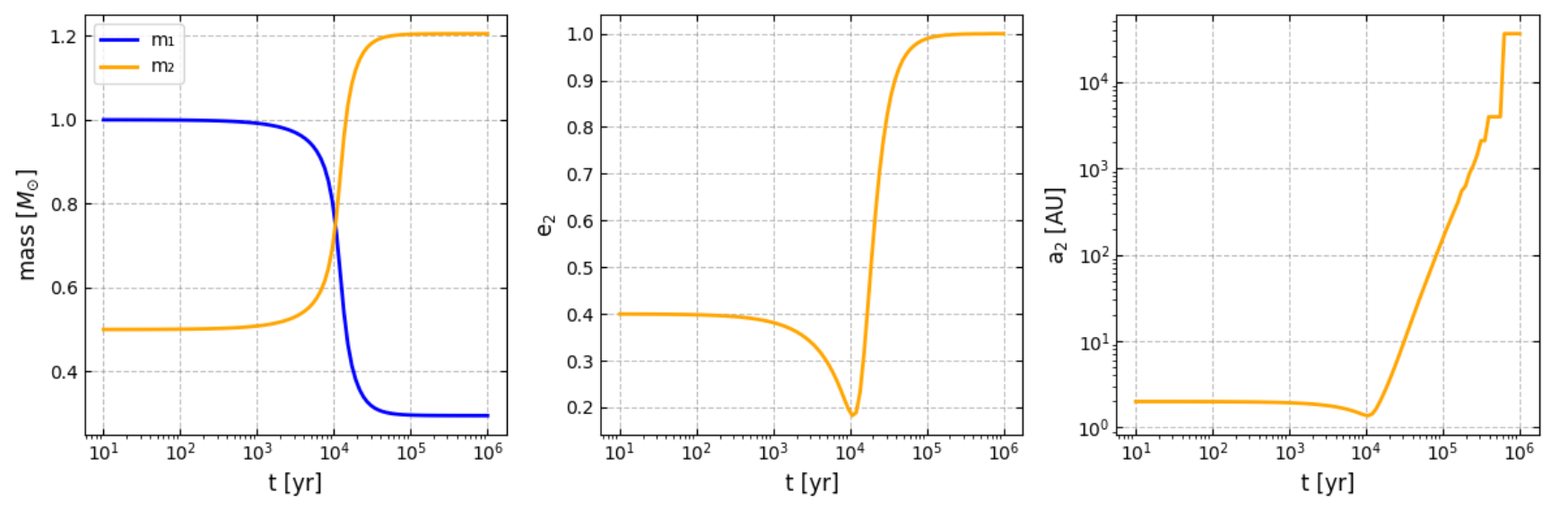}
    \caption{Left panel: The evolution of the binary components masses due to Roche lobe overflow. Center and right panels: the evolution of the secondary eccentricity and semimajor axis. }
    \label{fig:myplot3}
\end{figure}

In Figure \ref{fig:myplot3} we show an example for the evolution of a binary system due to the Roche lobe overflow.
{In this experiment we recover the characteristic two–phase eccentricity response expected for strongly phase‑dependent mass transfer in an eccentric binary. Because the overflow (here, effectively an exponential “leakage” channel given the chosen scale height) is concentrated near periastron, the orbital evolution is driven by a sequence of periastron‑centred impulses rather than a phase‑averaged secular torque, and the sign of the net eccentricity change depends sensitively on the instantaneous mass ratio. While the donor remains the more massive component, periastron‑peaked transfer tends to circularize the orbit: the mass and momentum redistribution (together with our minimal angular‑momentum consistency correction) preferentially reduces the osculating eccentricity, producing the initial dip. Once the system passes through mass‑ratio reversal, the same periastron‑localized transfer switches to an eccentricity‑exciting regime: the updates now act to increase the disparity between periastron and apastron motion, and, because our operator is non‑Hamiltonian and does not enforce orbital energy conservation even when the transfer is formally conservative, and because no tidal circularization is present to counteract excitation, the osculating eccentricity can grow secularly. In practice this \rev{leads} to a runaway toward very large eccentricities, with the orbital elements approaching the near‑parabolic limit as small cumulative changes in the energy–angular‑momentum balance are repeatedly deposited at the same orbital phase.}

\begin{lstlisting}[language=Python, caption={RLOF Python example}]
import math
import rebound
import reboundx
import numpy as np

# ---- Build a simple 2-body setup ----
sim = rebound.Simulation()
sim.units=['AU','yr','Msun']
(*@\textcolor{black}{sim.integrator = "ias15"}@*)
sim.dt= 0.1

sim.add(m=1.0, r=0.00465)     # donor (index 0)
sim.add(m=0.5, a=2., e=0.4, r=0.00465)  # accretor (index 1)

sim.move_to_com()

rx = reboundx.Extras(sim)
rlmt = rx.load_operator("roche_lobe_mass_transfer")
rx.add_operator(rlmt)

# Required operator params
rlmt.params["rlmt_donor"]     = 0  # donor is particle 0
rlmt.params["rlmt_accretor"]  = 1  # accretor is particle 1
rlmt.params["rlmt_skip_in_CE"] = 1

# RLOF controls
rlmt.params["rlmt_loss_fraction"] = 0.0
rlmt.params["jloss_mode"] = 0

sim.particles[0].params["rlmt_Hp"]    = 0.1
sim.particles[0].params["rlmt_mdot0"] = 1e-2

(*@\textcolor{black}{for t in np.logspace(1,6,100):}@*)
    sim.integrate((*@\textcolor{black}{t}@*))

\end{lstlisting}

\subsection*{Roche lobe overflow: Rebound vs MESA}
\begin{figure*}[h!]
    \centering
    \includegraphics[width=0.45\linewidth]{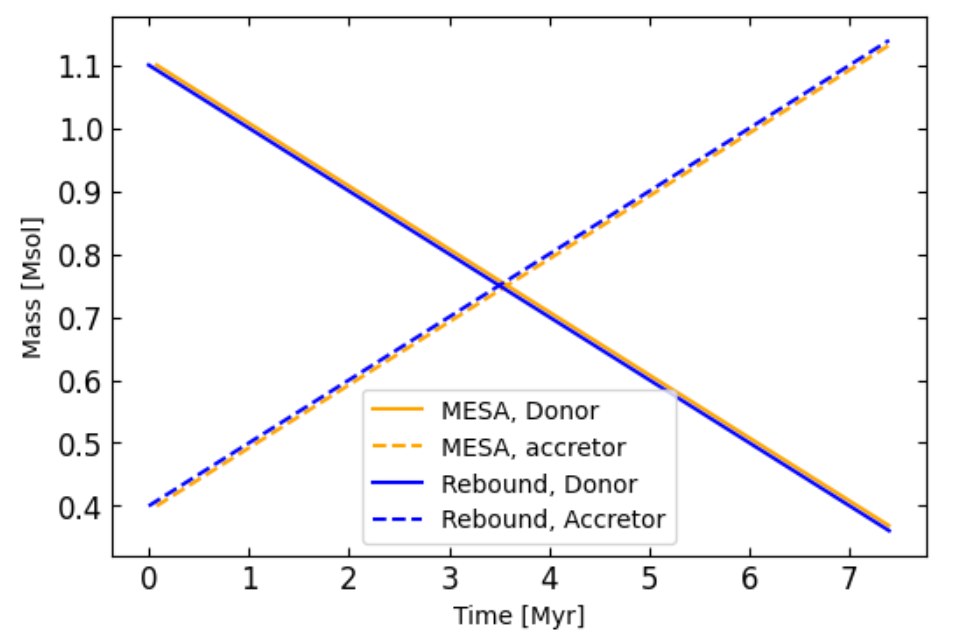}
    \includegraphics[width=0.45\linewidth]{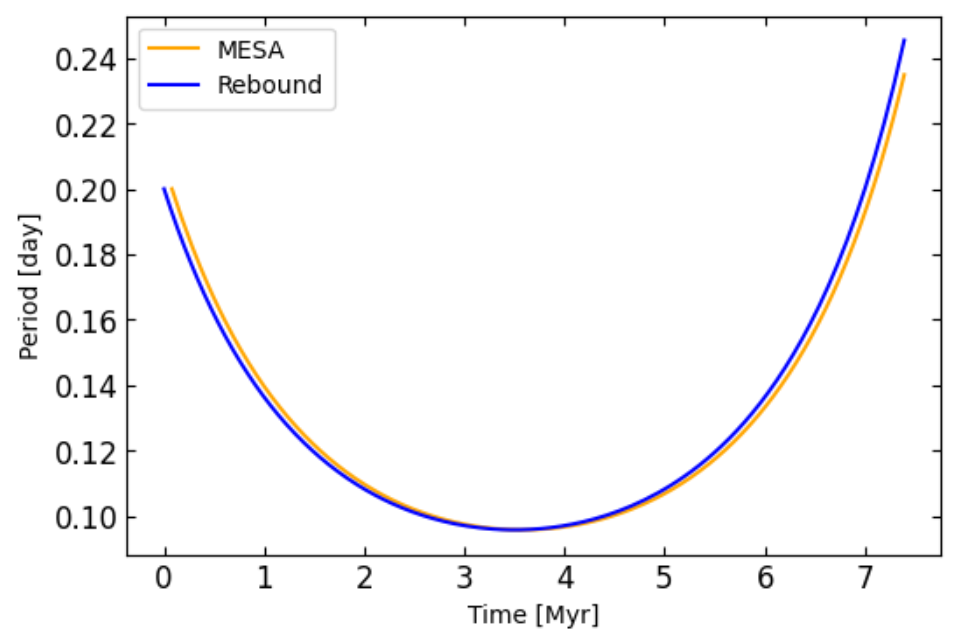}
    \caption{Left panel: time evolution of the donor and accretor masses for comparable Rebound and MESA models, due to RLOF.
Right panel: time evolution of the binary orbital period for the same case.}
    \label{fig:mesa1}
\end{figure*}

In Figure \ref{fig:mesa1} we show an example for the evolution of a binary system due to the Roche lobe overflow, compared to MESA.

\begin{lstlisting}[language=Python, caption={RLOF Python example for \textsc{REBOUNDx}--MESA comparison}]
import rebound
import reboundx
(*@\textcolor{black}{import numpy as np}@*)

# ---- Initiate the simulation ---- 
sim = rebound.Simulation() 
sim.units=['AU','yr','Msun'] 
(*@\textcolor{black}{sim.integrator = "ias15"}@*)

# Add donor and accretor
sim.add(m=1.1, r=0.00465)                          # donor (index 0)
sim.add(m=0.4, P=0.2 / 365.0, e=1e-4, r=0.00465)    # accretor (index 1)

# move the simulation to the center of mass
sim.move_to_com()

# -----------------------
# Roche lobe mass transfer
# -----------------------
rx = reboundx.Extras(sim)
rlmt = rx.load_operator("roche_lobe_mass_transfer")
rx.add_operator(rlmt)

# RLOF operator params 
rlmt.params["rlmt_donor"] = 0
rlmt.params["rlmt_accretor"] = 1
rlmt.params["rlmt_skip_in_CE"] = 1 # Skip common envelope evolution
rlmt.params["rlmt_loss_fraction"] = 0.0 # No mass loss to winds
rlmt.params["jloss_mode"] = 0
rlmt.params["rlmt_substep_max_dm"] = 0.1 # To speed up the simulation
rlmt.params["rlmt_substep_max_dr"] = 1000.0  # To speed up the simulation
rlmt.params["rlmt_min_substeps"] = 3 # default value

# Donor particle parameters (index 0)
# Hp is set to a large number to keep mdot = mdot0
sim.particles[0].params["rlmt_Hp"] = 10_000.0
sim.particles[0].params["rlmt_mdot0"] = 1e-7

# Integrate
for time in np.logspace(1, np.log10(7.39e6), 1000):
    sim.integrate(time)

\end{lstlisting}

\bibliographystyle{plainnat}

\end{document}